%


\input harvmac

\def\kpp{{K^{\prime\prime}}}
\def\kp{{K^\prime}}
\def\dGS{\delta_{\rm GS}}

%
\def\dslash{\not{\hbox{\kern-2pt $\partial$}}}
\def\Dslash{\not{\hbox{\kern-4pt $D$}}}
\def\Oslash{\not{\hbox{\kern-4pt $O$}}}
\def\Qslash{\not{\hbox{\kern-4pt $Q$}}}
\def\pslash{\not{\hbox{\kern-2.3pt $p$}}}
\def\kslash{\not{\hbox{\kern-2.3pt $k$}}}
\def\qslash{\not{\hbox{\kern-2.3pt $q$}}}
 \newtoks\slashfraction
 \slashfraction={.13}
 \def\slash#1{\setbox0\hbox{$ #1 $}
 \setbox0\hbox to \the\slashfraction\wd0{\hss \box0}/\box0 }


%

\noblackbox
\baselineskip 14pt plus 2pt minus 2pt
\Title{\vbox{\baselineskip12pt
\hbox{hep-th/9806124}
\hbox{SCIPP 98/20}
\hbox{WIS-98/15/June-DPP}
}}
{\vbox{
\centerline{Enhanced Symmetries and the Ground State}
\centerline{of String Theory}
  }}
\centerline{Michael Dine}
\medskip
\centerline{\it Santa Cruz Institute for Particle Physics}
\centerline{\it  University of California, Santa Cruz, CA 95064 }
\centerline{dine@scipp.ucsc.edu}
\bigskip
\centerline{ Yosef Nir and Yael Shadmi}
\medskip
\centerline{\it Department of Particle
Physics}
\centerline{\it Weizmann Institute of Science, Rehovot 76100, Israel}
\centerline{ftnir@wicc.weizmann.ac.il, yshadmi@wicc.weizmann.ac.il}
\bigskip

\baselineskip 18pt
\noindent
The ground state of string theory may
lie at a point of ``maximally enhanced symmetry", at which
all of the moduli transform under continuous or discrete
symmetries.  This hypothesis, along with the hypotheses
that the theory at high energies has $N=1$ supersymmetry
and that the gauge couplings are weak and unified, has definite consequences
for low energy physics.  We describe these, and offer some
suggestions as to how these assumptions might be compatible.

\Date{5/98}

\newsec{The Problem of Moduli}

%
\noindent
Moduli, exact and approximate, seem
almost ubiquitous in string ($M$) theory.  In recent
years, they have proven quite useful in elucidating
the underlying structure of the theory.
But they also pose one of the greatest challenges to the theory.
If string theory describes nature, how are they fixed?

The aim of this note is to enumerate,
in a very general way, mechanisms
by which this problem might be solved, and to
explore their consequences for low energy physics.
To study this problem, we will adopt a particular
framework.  We will assume, first, that the theory
has a fundamental energy scale comparable
to the Planck mass, $M_p$.  (This assumption
can be relaxed, but is convenient
for the initial discussion).  Slightly below this
scale, we will assume that the theory has $N=1$
supersymmetry, with some spectrum of light
particles, including the gauge
and matter fields of the MSSM.  We will assume that the
standard model gauge couplings at this scale
are small and unified, but that there is no weak
coupling description of the full theory.  Finally,
we will assume that supersymmetry breaking is
hierarchically small, i.e., it is of order $e^{-8 \pi^2/a g^2}$
for some constant $a$.

These are plausible hypotheses, but of course they need
not be true.  Adopting them, however, we can quickly
narrow down possible solutions to the moduli problem.
First, a definition:  we will use the phrase ``true vacuum
of string theory", to describe the putative state of the
theory which corresponds to the world we observe.
This does not mean that there are not many other
vacuum states; for example,  we expect there are
states with $N=2,4,8$ supersymmetry in four dimensions
which are exact non-perturbative solutions of the theory,
not to mention higher dimensional ground states with
varying degrees of supersymmetry.

There are a number of possibilities one can imagine for the true vacuum.  It
could be that this state contains approximate moduli.  By this we mean that
there are fields with potentials that tend to zero in some directions.
This definition includes states which have a weakly coupled
string theory limit or an $M$-theory limit.
Virtually all ideas about the moduli problem which have been discussed
to date involve approximate moduli.

Alternatively, it could be that in the true vacuum there are no approximate
moduli. One proposal along these lines was made in
\ref\dinesilverstein{M. Dine and E. Silverstein,
``New M Theory Vacua With Frozen Moduli," hep-th/9712166},
though that particular example did not have any supersymmetry.
If models exist without approximate moduli and unbroken supersymmetry at a high
 scale,\foot{
Such models presumably must be discovered in some self-consistent
scheme, since
by assumption supersymmetry is broken (albeit by a small amount)
and there is no small parameter.  However, it seems quite
possible that such theories exist.}
supersymmetry must be broken by low energy field theory
dynamics.\foot{Some related observations about theories without
moduli appear in \ref\burgess{C.P. Burgess, A. de la Macorra,
I. Maksymyk and F. Quevedo, ``Constant Versus Field
Dependent Gauge Couplings in Supersymmetric Theories,''
hep-ph/9805285.}.}  To understand
why this is so, one merely need note that (small) susy breaking
by high energy string effects must appear in the low
energy lagrangian through a linear superpotential for
some light fields; these fields are necessarily singlets under all of the 
symmetries. But by assumption, there
are no moduli whose $F$-terms can break supersymmetry. While it may be risky to
apply conventional notions of naturalness in string theory, it seems highly
unlikely that there are light fields which are singlets under all (gauge and
discrete) symmetries. So high energy effects cannot generate a supersymmetry
breaking superpotential.

Supersymmetry breaking, then, should be a phenomenon of the low energy 
field theory.  In other words,
the low energy field theory must contain a set
of fields and interactions which break supersymmetry\ref\trivedireview{E.
Poppitz and S. Trivedi, ``Dynamical
Supersymmetry Breaking,'' hep-th/9803107.}.  The scale of this breaking
is of order $Me^{-8 \pi^2/b g^2}$, where $g$ is the coupling
of this theory at the scale $M$ and $b$ is a constant.
According to our hypotheses
this coupling is small; note that it is not the expectation
value of a dynamical field.  It is important to stress that we
are not assuming that the string coupling is small or that the
complete theory is described by perturbation theory; only that there
is an effective low energy theory described by weak gauge couplings.
The plausibility of this assumption will be discussed further below.
Because there are no singlet fields, one cannot generate 
gaugino masses through non-renormalizable couplings to $W_{\alpha}^2$, 
as in conventional supergravity models.
So it is probably necessary to generate gaugino masses by gauge mediation.
Our hypotheses, then, lead almost inevitably to a picture of dynamical
(field theory) supersymmetry breaking and gauge mediation
\ref\ads{I. Affleck, M. Dine and N. Seiberg,
``Dynamical Supersymmetry Breaking and Its Phenomenological
Implications," Nucl. Phys. {\bf B256} (1985) 557.}
\ref\bkn{T. Banks, D. Kaplan and A.E. Nelson,
``Cosmological Implications of Dynamical Supersymmetry Breaking,"
Phys.~Rev. {\bf D49} (1994) 779.},
with the supersymmetry breaking scale anywhere between tens of TeV,
as in the models of
\ref\dns{M.~Dine, A.~Nelson, and Y.~Shirman, `` Dynamical Supersymmetry
 Breaking Simplified," Phys.~Rev. {\bf D51} (1995) 1362, hep-ph/9408384.}
\ref\dnns{M.~Dine, A.~Nelson, Y.~Nir, and Y.~Shirman, ``New Tools for
Low Energy Dynamical Supersymmetry Breaking," Phys.~Rev. {\bf D53}
(1996) 2658, hep-ph/9507378.},
and much higher scales, as in ``hybrid models" first discussed in
\ref\pt{E. Poppitz and S.P. Trivedi, ``New Models of Gauge and Gravity
 Mediated Supersymmetry Breaking," Phys.~Rev. {\bf D55} (1997) 5508,
 hep-ph/9609529.}.

Let us turn now to the question of theories with approximate moduli. 
It is usually assumed that the vacuum lies at some random point in 
the moduli space. But work on string dualities in the last few years 
suggests a more compelling possibility.  We now know that there are 
often points in the moduli space with ``maximally enhanced symmetry", 
i.e., points where all of the moduli are charged under some symmetry 
(which may be continuous or discrete).  For example, compactifying the 
heterotic string on $T_6$, there are many points where all of
the geometrical moduli are charged under enhanced gauge symmetries.  
Because these compactifications exhibit $S$-duality, at the self-dual 
point one has a $Z_2$ symmetry under which $S$ transforms.

This example is unrealistic, since it has $N=4$ supersymmetry, and
${g^2\over4\pi}=1$ at the self-dual point, which contradicts our
hypotheses.  But we now know of vast numbers of string compactifications
with high degrees of symmetry. It is
plausible that among these are states with maximally enhanced symmetry and
small, unified gauge couplings.  In general, maximally enhanced symmetry
points are of interest for at least two reasons:
\item{1.}  They are automatically stationary points of the effective
action, and thus candidate minima.  Tadpoles for the moduli are forbidden; 
it is a detailed question whether all of the moduli have positive masses.
\item{2.} They don't suffer from the cosmological moduli problem\bkn.
In thinking about string cosmology, it is usually assumed that the minimum of
the moduli potential lies at a random point in the moduli space, i.e., a point
with no special properties.  The universe is unlikely to start out at such a
point.  Finite temperature and/or curvature effects (say during inflation) are
likely to leave the universe somewhere else. Assuming the moduli potential is
of order the scale of supersymmetry breaking, the universe remains at this 
point until the Hubble parameter is of order the curvature of the potential, 
after which the system oscillates.  The moduli typically dominate the energy 
density of the universe when they decay, leading to catastrophic consequences. 
However, if the vacuum is a point of maximally enhanced symmetry, it is quite 
natural for the universe to start out in this state
\ref\drt{M. Dine, L. Randall and S. Thomas,
``Baryogenesis From Flat Directions of the Supersymmetric Standard
Model," Nucl. Phys. {\bf B458} (1996) 291, hep-ph/9507453.}.
For example, during inflation, even though the potential for the moduli
is modified, it quite naturally may prefer the symmetric state.
Finite temperature effects also tend
to prefer states of higher symmetry.  But even without any specific picture of
the universe, since maximally enhanced symmetry points {\it are} special, 
it is quite plausible that they will be singled out in the early universe.

Let us first explore the consequences of the assumption that the true vacuum
lies at an enhanced symmetry point.  Again, high energy string effects cannot 
be responsible for supersymmetry breaking, since the symmetries forbid any 
linear term in the moduli superpotential.  In addition, gluino condensation 
cannot yield supersymmetry breaking in such a  vacuum, since couplings
${\cal M} W_\alpha^2$ are forbidden.
So the situation is very similar to that in theories without moduli;
even if the moduli acquire substantial $F$-components, because of the
symmetries they cannot couple to $W_{\alpha}^2$,
and thus cannot contribute to gaugino masses.
Note  that field theories with dynamical supersymmetry
breaking similarly do not contain fields that are singlets under all
symmetries\foot{Models such as those of Intriligator and Thomas
\ref\intthomas{K.~Intriligator and S.~Thomas,
``Dynamical Supersymmetry Breaking on Quantum Moduli
Spaces," Nucl. Phys. {\bf B473} (1996) 121, hep-th/9603158.}\
or Nelson
\ref\nelson{A.E.~Nelson,
``A Viable Model of Dynamical of Supersymmetry Breaking in the Hidden Sector,"
Phys. Lett. {\bf B369} (1996) 277, hep-ph/9511350.}
are only natural if the various singlets transform under discrete symmetries.}.
So, again, we are led to a picture of dynamical supersymmetry breaking
in the effective low-energy field theory, almost surely involving some
sort of hidden sector and gauge
mediation.

How would the moduli be fixed in such a picture?  One can imagine several
possibilities.  The first question, though, is to identify the moduli.  At the
renormalizable level, the MSSM possesses several flat directions \drt.  Some of
these could, in fact, be exact at the level of the superpotential. For example,
there is a set of flat directions labeled by chiral fields $\bar u\bar d \bar
d$. These would be exactly flat if one had a (discrete) R symmetry under which
the $\bar u$, $\bar d$ and $\bar e$ fields were neutral. It is also possible to
find flat directions for which all of the standard model gauge symmetries are
broken. (One such direction has non-zero vev's for the $Q$ and $L$ fields, and
discrete symmetries; these directions are described by $12$ parameters, i.e.,
they could correspond to as many as $12$ moduli.) So it could be that the 
moduli are in fact linear combinations of the ordinary squarks, sleptons and 
Higgs fields. The enhanced symmetries might just be the symmetries of the 
standard model!

If this is the case, it is easy to imagine how the moduli are fixed. In gauge
mediation, for example, the squarks and sleptons gain mass from K\"ahler
potential corrections; for a sensible model, the curvature of the potential
must be positive at the origin, i.e., at the enhanced symmetry point.
This is, indeed, a realization of the ``K\"ahler stabilization"
of~\ref\coping{T. Banks and M. Dine,
``Coping with Strongly Coupled String Theory,"
Phys. Rev. {\bf D50} (1994) 7454, hep-th/9406132.}.

Alternatively, some or all of the moduli might
play a role in the hidden sector dynamics.
The models of \intthomas\ and \nelson, for example,
have classical moduli.  These moduli may be
stabilized near the origin, i.e., near the
enhanced symmetry point.  In this case, the masses
of the moduli could easily be $10$'s of TeV or larger.

Finally, some or all of the moduli might
be singlets both with respect to the standard
model gauge group and with respect to any hidden
sector gauge group.  Assuming that supersymmetry
is broken by the $F$-components of some hidden sector
fields, $Z$,  these fields could gain mass
through K\"ahler potential terms such as
\eqn\singletmoduli{K = {\cal M}^{\dagger}{\cal M}Z^{\dagger}Z\ .}
Such couplings respect the symmetries,
and if the sign is suitable, stabilize the
moduli at the symmetry point.  The moduli fields would be quite light.
However, because of the symmetries, their interactions with
ordinary matter would be highly suppressed.  For example,
typical axion or Brans-Dicke couplings would be forbidden.
Couplings to quarks and leptons would also be suppressed by
extra powers of the large mass scale, the amount of
suppression depending on the charges of the quarks and leptons.

\newsec{The Cosmological Constant Problem}
%
\noindent
One obvious question in this picture is:  can the cosmological constant vanish?
We have no new wisdom to offer as to the detailed mechanisms by which the
cosmological constant vanishes.  In particular, we do not know whether the
vacuum energy should vanish or nearly vanish in the effective action. Still, we
can give conditions under which cancellation of the vacuum energy at the level
of the effective action is at least possible. In conventional supergravity model
building, one starts with the potential
\eqn\sugra{V=e^{K}\left[\left(
{\partial W\over\partial\phi_i}+{\partial K\over
\partial\phi_i}W\right)K^{i\bar i}
\left({\partial W^*\over\partial\phi_{\bar i}}
+{\partial K\over\partial\phi_{\bar i}}W^*\right)-3\vert W\vert^2\right]\ .}
One then notes that the superpotential can include a constant, and this
constant is adjusted to give zero potential at the minimum.

We wish, then, to ask whether there can be a constant of the correct order of
magnitude, in the context of models
with maximally enhanced symmetry.
Consider, first, theories with no moduli.  In such theories, one
would expect that $W \sim 1$, unless there is a symmetry.  This is presumably a
disaster, but such a term can be forbidden by a discrete $R$ symmetry. In order
that there be a small constant term in $W$, one then requires a spontaneous
breaking of the discrete symmetry in the low energy theory.  This can occur, for
example, if there is a sector containing a pure gauge theory without matter. In
the absence of moduli, such a sector can generate a non-zero, constant value for
$W$.  This can be of the correct order of magnitude (this depends on the
$\beta$-functions of the various sectors).

In the case where there are approximate moduli, there are similar issues. It is
possible, as we will describe below, that $e^{-S}$ is numerically small.
Alternatively, it may be forbidden by discrete symmetries.  However, in either
case, obtaining a constant of the correct order of magnitude may require a
sector like that described above to generate the necessary constant. Of course,
in any of these pictures, the precise cancellation of the cosmological constant
looks miraculous from the perspective of the effective field theory.

\newsec{Why are the couplings small and unified?}
%
\noindent
One of the great puzzles of string theory is:  why are the gauge
couplings we observe small and unified, if the underlying theory is
strongly coupled.  The weakly coupled heterotic
string provides a simple picture of coupling unification, but it
is not clear why this should hold at stronger coupling, when
$e^{-S}$ cannot be neglected. The picture
we have developed of maximally enhanced symmetries provides,
perhaps, a plausible answer.  At such points, the gauge couplings
will be pure numbers, independent of any moduli, determined,
perhaps, by topological considerations.  One can
certainly conjecture that these numbers are sometimes small,
and occasionally equal.  This seems more plausible than equality
in theories where the string coupling
is large but the moduli are simply random.

Alternatively, in \coping, a scenario for coupling unification at
small gauge coupling was proposed, along with a mechanism for
stabilizing the dilaton. It was argued that perhaps $e^{-S}$ is small,
so there are no large corrections to the superpotential, $W$,
and the gauge coupling function, $f$, but that corrections
to the K\"ahler potential might be large,
and that these might be responsible for the stabilization of the dilaton.
The point is that while one expects that string perturbation theory
is less convergent than the perturbation expansion of field theory
\ref\shenker{S. Shenker, talk at Cargese workshop on Random
 Surfaces, Quantum Gravity and Strings, France (1990).},
holomorphy severely restricts the form of the corrections
to $W$ and $f$.

We can ask whether we can account for weak coupling
and unification in a similar way in the case of maximally
enhanced symmetry.  In the simple example of maximally
enhanced symmetry discussed in the introduction, ${g^2\over4\pi}$
is equal to one.  Indeed, we might expect this behavior is generic.
Consider starting with some weak coupling description, and varying
the dilaton.  At weak coupling, the superpotential is independent
of $S$, up to terms of order $e^{-S}$.   However, if a
symmetry is to be restored, we would expect that the corrections
to the superpotential must be large, so that couplings which violate
the symmetry vanish,
and perhaps that massive particles become massless.
This requires that $e^{-S}$ not be negligible at the enhanced
symmetry point.  In such a view, no matter what the numerical
value of the coupling, there would be no sense in which one
could treat $e^{-{8 \pi^2 \over g^2}}$ as small
in the full string theory, and coupling unification would have
to be a consequence of a particular choice of strongly coupled
ground state.

The possible loophole to this argument is, again, the K\"ahler potential.
Could the K\"ahler potential be responsible for symmetry restoration?
If this is to be the case, $K$
presumably must be singular at the symmetry point.
While we have no theory where we can currently
exhibit such an effect in a meaningful calculation, in the
following section we offer a toy model for this phenomenon.

\newsec{A Model for Symmetry Enhancement Through K\"ahler
Potential Effects}
%
\noindent
In \coping, in a weak coupling language,
it was suggested that perhaps the moduli are stabilized
in a region where corrections to the classical K\"ahler potential
are large, but $e^{-S}$ is very small.
Based on the argument above, we would not expect enhanced
symmetries at such a point, {\it unless} the K\"ahler potential
were singular.\foot{Ref. \coping\ actually argued in the other direction:
that symmetries which are unbroken
at weak coupling can only be broken at stronger coupling if the
K\"ahler potential becomes singular.}  The question is:
what sort of singularities are required and how
plausible is such singular behavior for the K\"ahler potential?

Ref.\ref\bdtwo{T. Banks and M. Dine,
``Quantum Moduli Spaces of N=1 String Theories,"
Phys.Rev. {\bf D53} (1996) 5790, hep-th/9508071.}\
discussed a possibility for
stabilizing the dilaton involving Fayet-Iliopoulos terms.  It was argued that
if the $D$-term vanishes somewhere on the moduli space, it is natural that the
potential has a minimum nearby. A little thought indicates that a vanishing
$D$-term also provides a model for enhanced symmetries. In models where a gauge
anomaly is cancelled by the axion, there is a Fayet-Iliopoulos $D$-term, which,
for general points on the moduli space, is given by
\eqn\fidterm{\xi^2= -{\dGS\over 2}\, {\partial K \over \partial S}\ ,}
where
\eqn\deltags{\dGS = {1 \over 192 \pi^2}\sum_i q_i\ .}
At generic points in the moduli space, the Green-Schwarz term
is non-zero; typically the $D$-term is cancelled by the
expectation value of some scalar field(s), $N$.  This scalar field expectation
value may break discrete or continuous symmetries in addition
to the anomalous $U(1)$.  Now
if the derivative were to vanish at some point in the moduli space,
the expectation value would vanish, and symmetries might
be restored.  This symmetry restoration is controlled by
a non-holomorphic quantity. So there is the possibility of enhanced
symmetry, and small holomorphic coupling.

But, as presented, there is a problem with this idea.
In order that the derivative of the K\"ahler potential vanish,
it is necessary that the second derivative, ${\partial^2 K\over
\partial S^2}$, change sign.
This means, however, that the kinetic term for the dilaton
does not have the correct sign.  In \bdtwo, it was asserted
that this would not be a problem once kinetic mixing with
other moduli was considered.  But this is not the case; it is
easy to show that for a general symmetric matrix, if a diagonal
component of the matrix is negative, at least one eigenvalue
must be negative. So this description of the problem cannot be correct.

However, there is no reason why the description
of the theory in terms of various light fields, the dilaton
multiplet, and the massive $U(1)$ gauge multiplet, should
make sense for such values of the coupling.  At such points,
the gauge multiplet has mass of order the other heavy fields
in the theory.  It also has a very large width.  So it is
not clear that we should be keeping only this set of degrees
of freedom. Instead, we should focus on the light degrees of freedom,
and ask what features of the K\"ahler potential are responsible
for symmetry restoration.

It will prove instructive to first consider the $D$-term model
at weak coupling. Loosely speaking, this limit already  provides 
us with an example of symmetry restoration through K\"ahler 
potential effects. Let us suppose that
there is a single field, $N$, with charge $q_N<0$ (we take $\dGS>0$), which
cancels the $D$-term (in particular, the direction with non-zero
$N$ is ``$F$-flat'').
The weakly coupled theory is described by a K\"ahler potential
\eqn\kahler{K= f(S+ S^{\dagger} - \dGS V)+ N^{\dagger}
e^{2q_NV}N + \sum \phi_i^{\dagger}e^{2q_i V}\phi_i + \dots\ ,}
where $\phi_i$ are light, chiral fields with $U(1)$ charge $q_i$,
which, we assume, do not develop vevs.
The $D$-term is given in eqn. \fidterm, so
\eqn\nvev{N^{\dagger}N= {\dGS \over 2 q_N}\, {\partial K \over \partial S}\ .}
Since the $N$ field is massive (the light modulus at weak coupling
is principally $S$), it is appropriate to integrate it out in this region.
The theory will also contain, in general, superpotential terms
such as (suppressing couplings)
\eqn\samplew{W = N \phi_i \phi_j + N \phi_i \phi_j \phi_k + \dots\ .}
The expectation value of $N$ then leads to masses, Yukawa couplings,
and so on in the low energy theory.  Because the $N$ expectation
value is a {\it non-holomorphic} function of the dilaton,
so are these couplings. Moreover, the $N$ contribution to the masses
leads to non-holomorphic corrections to the gauge couplings.
Taking $S \to \infty$, the $N$-vev tends to zero, so that various masses and
couplings in eqn.~\samplew\ approach zero as well.
Thus, even at weak coupling we have a model of ``symmetry restoration".
In the low energy theory, it must be possible to interpret the
$S$-dependence of the couplings in terms of the K\"ahler
potential for the light fields.

In order to understand the structure of the low energy theory,
first note that the heavy scalar field, which is eaten
by the $U(1)$ gauge field, is of the form,
$H \sim \dGS \kpp\,  S - 2q_N \vert\langle N\rangle\vert^2 \, \ln N$,
and is predominantly $N$,
while the light field, which parametrizes the flat direction, is the
gauge-invariant combination
$ S + {\dGS\over 2 q_N} \ln N$, which, at weak coupling, is approximately $S$.
It is therefore convenient to choose a  gauge in which $N$ is
fixed, $N=n_0$, for some fixed reference $n_0$.
Note that this is a holomorphic gauge condition.
In this gauge the K\"ahler potential can be written as
\eqn\kahlern{K= f(\Phi+ \Phi^{\dagger} - \dGS X)+
 \vert n_0\vert^2 e^{2q_N X} + \sum \phi_i^{\dagger}e^{2q_i X}\phi_i
+ \dots\ ,}
where $\Phi = S + {\dGS\over 2 q_N} \ln {N\over n_0}$.
Integrating out the heavy field $X$ we have
\eqn\eom{
2 q_N \vert n_0\vert^2 e^{2 q_N X} - \dGS
\kp(\Phi + \Phi^\dagger -\dGS X) = 0\ ,
}
so that, for large $\Phi$,
$X \sim {1\over 2 q_N  \vert n_0\vert^2}
\ln(\left\vert{\dGS\over 2 q_N} \right\vert {1\over \Phi+\Phi^\dagger})$.
At low energy we then have the fields $\Phi$, $\phi_i$
with the K\"ahler potential
\eqn\kahlerlow{K= -\ln(\Phi+ \Phi^{\dagger})+
 \sum Z_i\phi_i^{\dagger}\phi_i + \dots\ ,}
with
\eqn\zi{Z_i =
\left(\left\vert{\dGS\over 2 q_N  } \right\vert \,
{1\over \vert n_0\vert^2}\,
{1\over \Phi+\Phi^\dagger}\right)^{q_i\over q_N} \ =
\left\vert {{\langle N \rangle\over n_0}}\right\vert^{2q_i\over q_N}\ .
}
\noindent From the point of view of this low energy theory,
(which does not have $N$ in the superpotential), symmetry
restoration occurs since the K\"ahler potential
becomes singular as $\Phi \to \infty$.
Note that the singularity is just a consequence of
our insistence of writing the theory in a holomorphic fashion.
Physical quantities are not singular.

The question, then, is whether
such singularities can occur on the ``interior'' of the
moduli space, as we vary the gauge coupling.
We have seen that the notion of a vanishing $D$-term at an
interior point is, at best, a caricature of what might be going
on; because of the change of sign of the kinetic term, such
a description cannot be complete.  Still, it is a suggestive
model.  Suppose, for a moment, that it is a suitable description
near the enhanced symmetry point. We could imagine repeating the 
previous exercise near the point $\kp =0$.
In order to smoothly interpolate between the weak and strong coupling
regions, it is again useful to choose the gauge $N=n_0$.  From a physical
point of view, this is a rather peculiar choice, however, as the
modulus, $\Phi$, in this regime is predominantly $N$.  It is helpful,
indeed, to consider first the gauge $S= s_0$, and then gauge
transform to the gauge with constant $N$, through the gauge transformation
$i\Lambda = \ln (N/n_o)$.
Proceeding in this way, near the enhanced symmetry point $\langle N\rangle=0$,
\eqn\zizeron{
Z_i = {\left({N\over n_0}\right)}^{q_i/q_N}
{{\left({N\over n_0}\right)}^\dagger}^{q_i/q_N} =
 \left( {(e^{\Phi-\Phi_0})}^\dagger e^{\Phi-\Phi_0}\right)^{2q_i\over\dGS} \ ,
}
where $\Phi_0$ is just the constant $s_0$.
As expected, the $Z_i$
become singular as $ N\to 0$ (or $\Phi \to \infty$),
leading to symmetry restoration.
Note that in this gauge, at the enhanced symmetry point,
the $Z$'s are  products of a holomorphic and
an antiholomorphic factor
and so we can make a {\it holomorphic} redefinition
of the fields to remove the $Z_i$'s from the K\"ahler
potential.  This redefinition puts $N$, now interpreted
as a light, dynamical field, in the appropriate places
in the superpotential. Indeed, this redefinition is just the
gauge transformation in reverse, so we recover the 
$S=s_o$-gauge description.
One should point out that in this model, there are not necessarily
new massless particles at the enhanced symmetry point.
If there are, one can contemplate more complicated singularities.

It is interesting to note in this description how the gauge coupling function
respects the symmetry.  After all, one of the main points in our
discussion of enhanced symmetries is that there cannot be couplings
of the charged modulus to the gauge fields,
or linear terms in the charged modulus in the superpotential.
In the gauge $N=n_0$, the gauge coupling function is equal to
$\Phi$, which does transform under the symmetry.
However, the redefinitions of
the $\phi_i$'s, required to make their kinetic terms canonical,
precisely cancel the $N$-dependent term in $\Phi$,
leaving $s_0$ in front of the gauge kinetic terms.

We can thus describe the phenomena which occur in this model
entirely in terms of the K\"ahler potential of the light
fields. The crucial features are that at the enhanced
symmetry point, the K\"ahler potential for the modulus, $\Phi$,
behaves as
\eqn\kphinew{K_\Phi = \vert n_0\vert^2
 \left( {(e^{\Phi-\Phi_0})}^\dagger e^{\Phi-\Phi_0}\right)^{2q_N\over\dGS}\ ,
}
and $Z_i(\Phi)$ is given in eqn.~\zizeron,
whereas at weak coupling $K_\Phi$ and $Z_i(\Phi)$ are
as given in eqns.~\kahlerlow-\zi.
This example makes clear that K\"ahler potential effects
can, in principle, restore a symmetry.
While phrasing the discussion in terms of the light
modulus $\Phi$ seems to avoid the issue of  $\kpp < 0$,
it should be noted that the K\"ahler potential for $\Phi$
is not single-valued;
$\Phi \to \infty$ both in the weak-coupling limit and
in the limit that the symmetry is exactly restored.
We do not know how to assess the plausibility of such behavior of the
K\"ahler potential in real string theory.

Another toy example of symmetry restoration through K\"ahler
effects, that does not involve  a negative kinetic term,
has two anomalous $U(1)$'s. Several anomalous $U(1)$'s can appear
in non-perturbative heterotic vacua (or in perturbative type I vacua)
where there can be more than one dilaton field to cancel
the $U(1)$ anomalies
\ref\AFIV{G. Aldazabal, A. Font, L.E. Ibanez and G. Violero,
 ``D=4, N=1, Type IIB Orientifolds," hep-th/9804026.}
\ref\Iban{L.E. Ibanez, ``New Perspectives in String Phenomenology
 from Dualities," hep-ph/9804236.}.
Consider then a model in which the Fayet-Iliopoulos terms are canceled,
at generic points, by the vevs of two fields,
$N_1$ and $N_2$, of charges $(q_\alpha^a, q_\alpha^b)$, $\alpha=1,2$, under
$U(1)_a$, $U(1)_b$. It is easy to see that the $N_1$ vev vanishes for
\eqn\ZeroVev{
q_2^b\dGS^a{\partial K\over\partial S_a}=
q_2^a\dGS^b{\partial K\over\partial S_b}\ .
}
If $N_1$ transforms under an additional symmetry,
this symmetry is restored along the curve~\ZeroVev.
Note that such a curve exists even if  $K$ is
perturbative.
(We do not know whether this is realized in actual
$F$-theory examples.)
Thus there may be a region of moduli space that contains curves of enhanced
symmetry, but no negative kinetic terms or branched behavior.
Here again, in the low energy theory, the enhanced symmetry is associated with
singular wave-function renormalizations for some of the light fields.
This model, however,
does not lead to {\it maximal} symmetry restoration
along $N_1 =0, N_2\neq 0$.

\newsec{Conclusions}
%
\noindent
In considering the problem of supersymmetry breaking
in string theory, it has almost invariably been
assumed that the true string vacuum contains approximate
moduli, and that they are fixed at some random
value.  In this note, we have considered two alternative
possibilities:  either the vacuum has no approximate
moduli, or the moduli sit at or very near a point
of maximally enhanced symmetry.  In both cases,
we have argued that supersymmetry breaking must occur
in the low-energy field theory, and that superpartner masses
are generated, at least in part, by gauge mediation.

In the latter case, we have pointed out that the
moduli might actually be the superpartners of
ordinary quarks and leptons.  The same mechanisms
which give them mass (e.g. gauge mediation) then also
fix their expectation values at the symmetry point.
Alternatively, the moduli might be very light, and
extremely weakly coupled, or they might be fields
which participate in a supersymmetry-breaking hidden sector.

We have noted that the moduli problem of cosmology
is readily solved in such a framework.  This is simply
because the zero energy ground state is a special
point.  Finite temperature or curvature effects
can naturally favor such enhanced symmetry points.

We have not seen a clear, low energy signal which
distinguishes between the no moduli/approximate
moduli alternatives.  Since the MSSM contains
approximate flat directions, only measurements
of high dimension operators determine whether the
directions are truly flat.
One could imagine, however, a string theory program
in which one would enumerate, e.g. $F$-theory
compactifications with suitable gauge groups, particle
content, symmetries and couplings
\ref\KaLo{V. Kaplunovsky and J. Louis, ``Phenomenological
Aspects of F-Theory," Phys. Lett. {\bf B417} (1998) 45, hep-th/9708049.}.
Choosing such a state would then determine the low energy gauge group
and the detailed pattern of supersymmetry breaking.

It should be noted that it is possible to relax a number of our assumptions.
For example, one could imagine that there are a variety of scales, as in
\ref\wittenm{E. Witten, ``Strong Coupling Expansion of Calabi-Yau
Compactification," Nucl. Phys. {\bf B471} (1996) 135, hep-th/9602070.}
\ref\bdm{T. Banks and M. Dine, ``Couplings and Scales in
Strongly Coupled String Theory,"
Nucl. Phys. {\bf B479} (1996) 173, hep-th/9605136.}
\ref\horava{P. Horava, ``Gluino Condensation in Strongly
Coupled Heterotic String Theory,"
Phys. Rev. {\bf D54} (1996) 7561, hep-th/9608019.}
\ref\dimopoulosetal{N. Arkani-Hamed, S. Dimopoulos and G. Dvali,
``The Hierarchy Problem and New Dimensions at a Millimeter," hep-ph/9803315.}.
In these cases, one might want to understand not only why the gauge couplings
are small, but also why certain ``geometrical moduli" are also large or small.
This might be relevant to understanding flavor structure
(as might be the enhanced symmetries).

In any case, we believe that these ideas establish a
framework for considering the problem of the moduli,
which can be pursued both at the level of phenomenological
model building and of more fundamental string theory.

\bigskip
\noindent
{\bf Acknowledgements}
\smallskip
\noindent
We thank David Kutasov, Jan Louis, Eva Silverstein, Nathan Seiberg,
Adam Schwimmer and Scott Thomas for conversations.  We are especially
grateful to Tom Banks for critical comments on an early version of
this manuscript. This work was supported in part by the
United States -- Israel Binational Science Foundation (BSF).
The work of M.D. was supported in part by the U.S. Department
of Energy.  Y.N. is supported in part by the Israel Science
Foundation and by the Minerva Foundation (Munich).

\listrefs

\bye